\documentclass[twocolumn,showpacs,preprintnumbers,prl,amsfonts,amsmath,amssymb]{revtex4-1}

\usepackage{graphicx}
\usepackage{dcolumn}
\usepackage{bbm}
\usepackage{bm}

\newcommand{ \mo }[1]{ \hat{ #1 } }

\newcommand{ \mv }[1]{ \mathbf{ #1 } }

\newcommand{ \sH }{ \mathcal{H} }

\newcommand{ \Tr }{ \mbox{Tr} }

\newcommand{\be}{\begin{equation}}
\newcommand{\ee}{\end{equation}}
\newcommand{\ba}{\begin{array}}
\newcommand{\ea}{\end{array}}
\newcommand{\bea}{\begin{eqnarray}}
\newcommand{\eea}{\end{eqnarray}}

\newcommand{ \sEf }[1]{ \mathbb{E}_f\left[ #1 \right] }
\newcommand{ \sEq }[2]{ \mathbb{E}_{#1}\left[ #2 \right] }

\begin{document}

\title{Number-Phase Wigner Representation for Scalable Stochastic Simulations of Controlled Quantum Systems}

\author{M. R. Hush}
\author{A. R. R. Carvalho}
\author{J. J. Hope}
\affiliation{Department of Quantum Science, Research School of Physics and Engineering, The Australian National University, ACT 0200, Australia}

\date{\today}

\begin{abstract}



Simulation of conditional master equations is important to describe systems under continuous measurement and for the design of control strategies in quantum systems. For large bosonic systems, such as BEC and atom lasers, full quantum field simulations must rely on scalable stochastic methods whose convergence time is restricted by the use of representations based on coherent states. Here we show that typical measurements on atom-optical systems have a common form that allows for an efficient simulation using the number-phase Wigner (NPW) phase-space representation. We demonstrate that a stochastic method based on the NPW can converge over an order of magnitude longer and more precisely than its coherent equivalent. This opens the possibility of realistic simulations of controlled multi-mode quantum systems.

\end{abstract}


\maketitle

Exciting advances in physics have led to a boom of research into technologies that exploit fundamental quantum properties. Such quantum technologies now encompass more than lasers and superconductors. Indeed, there are applications to precision metrology \cite{Giovannetti:2006,Leibfried:2004}, quantum information processing and quantum cryptography \cite{Nielsen:2000,Gisin:2002}. A key feature of quantum technologies is that they require the precise creation, measurement and control of individual quantum systems. In particular, measurement-based feedback control has shown promise as an effective and robust technique for controlling quantum systems. 
The first experiments \cite{Mabuchi:2002a,Steck:2004} and much theoretical work \cite{Belavkin:1983,Wiseman:1993,Doherty:1999,Ramon-van-Handel:2005,Szigeti:2009,Szigeti:2010} on feedback control of quantum systems have been applied to relatively low-dimensional systems. This letter describes a technique for efficient simulation of large bosonic conditional quantum systems that is more than an order of magnitude more precise and converges for significantly longer timescales than previous methods, and that scales logarithmically with the size of the Hilbert space.

A large bosonic system of particular interest to quantum science is the Bose-Einstein condensate. Measurement-based feedback control of BECs and atom lasers was first investigated in a single-mode model, where a continuous number measurement was used to reduce the interaction-induced phase diffusion that limits single-mode atom laser linewidth \cite{Wiseman:2001}.  It was then shown that position measurement and feedback on a single trapped atom could bring it to the ground state \cite{Doherty:1999}, but the proposed measurement scheme was not suitable for large atomic clouds such as a condensate.  A multimode quantum field model of a condensate measured by an existing experimental technique (phase-contrast imaging) was then produced, but it could only be solved using a semiclassical approximation \cite{Szigeti:2009,Szigeti:2010}.  Analysis of the linewidth of a multimode atom laser undergoing feedback will require a viable stochastic method for conditional quantum states that can deal with both high nonlinearities and number-like measurements.  The number-phase Wigner function method fulfills both of these requirements. 

The most effective methods for dynamic simulation of high-dimensional bosonic quantum systems are stochastic techniques based on phase-space representations \cite{Gardiner:1983,Gardiner:1991}. Each stochastic method is derived from a specific phase-space representation, which is akin to the choice of a basis for the Hilbert space. Naively, these techniques require memory and computational resources that scale logarithmically with the size of the Hilbert space. Practically, the overall computational efficiency is system dependent, and strongly depends on how well the underlying phase-space representation matches the natural basis for the quantum system under consideration. The most commonly used stochastic simulation methods are based on phase-space representations that use Gaussian states. These methods have enabled the simulation of quantum optical \cite{Gardiner:1991,Drummond:1987}, atomic \cite{Drummond:1999,Wiseman:2001,Hope:2001}, and fermionic quantum fields \cite{Corney:2004}. In particular, stochastic methods have been used extensively in the field of quantum-atom optics, where dilute atomic gases can be cooled to produce BECs and atom lasers \cite{Steel:1998, Dall:2009, Savage:2007}.  The two most successful varieties are based on positive P (P$^+$) and truncated Wigner (TW) representations. P$^+$ is an exact technique, but requires a doubling of the phase space that often leads to instabilities \cite{Gilchrist:1997}. Truncated Wigner is an approximate technique that typically has significantly longer convergence times than P$^+$. However it makes an uncontrolled approximation \cite{Sinatra:2002}, and may therefore converge to incorrect solutions.  Both of these methods, along with all other coherent-state based representations, experience difficulties dealing with large number-conserving nonlinearities, as the underlying Gaussian basis becomes inappropriate. Such large number-conserving nonlinearities are typically the dominant energies in confined cold atomic systems by a couple of orders of magnitude. Recently, we introduced a new stochastic method based on a number-phase Wigner (NPW) representation \cite{Hush:2010}, that provides a non-approximate method for simulating large number-conserving nonlinearities. It was found that this dramatically improved the convergence of simulations of these highly nonlinear systems.   

Modelling highly nonlinear systems undergoing continuous monitoring and feedback requires the simulation of a \emph{conditional} quantum state.  In a recent paper \cite{Hush:2009}, we extended stochastic simulation techniques to apply to a class of conditional quantum systems. Continuous measurement of a quantum system can have a dramatic effect on its dynamics.  In fact the choice of measurement can even be used as a controlling mechanism by itself \cite{Jacobs:2009}.  It is therefore unsurprising that the appropriate choice of phase-space representation is heavily influenced by the choice of measurement, as it may drive the conditioned system towards a state that is simpler to describe in a particular representation.  Also, not all measurements automatically produce a usable method.  Our previous paper demonstrated how to unravel a particular form of stochastic Fokker-Planck equation (SFPE) \cite{Hush:2009}, and it is only possible to generate SFPEs of this form with particular combinations of measurement schemes and phase-space representations. In particular, methods based on coherent state representations are badly suited to measurements involving number-like observables rather than quadrature-like observables, which are prevalent in atom optics.  In this letter, we show that the NPW representation produces dramatically superior results than coherent state based representations for these calculations.

A common quantum atom-optical system under monitoring is governed by the conditional master equation
\begin{equation}
d\mo{\rho} = -i[\mo{H},\mo{\rho}] \; dt + \sum_i \mathcal{D}[\mo{L}_i] \mo{\rho} \; dt + \sum_i \mathcal{H}[\mo{L}_i] \mo{\rho} \; dW_i.  
\label{eqn:mastmulti}
\end{equation}
where $dW$ is an Ito Wiener increment; $\mathcal{D}[\mo{c}] \mo{\rho} = \mo{c}\mo{\rho}\mo{c}^\dag - \frac{1}{2}(\mo{c}^\dag\mo{c}\rho + \rho\mo{c}^\dag\mo{c})$; $\sH[\mo{c}] = \mo{c}\rho + \rho\mo{c}^\dag - \Tr[ \mo{c} \rho + \rho \mo{c}^\dag] \mo{\rho} $; $\mo{H}$ is the Hamiltonian and contains the contributions from kinetic, potential and many-body interaction energies; and $\mo{L}_i = \int d\mv{x} \mo{\psi}^\dag(\mv{x}) L_i(\mv{x}) \mo{\psi}(\mv{x}) $ is the measurement operator where  $L_i(\mv{x}) \in \mathbf{L^2}$ are the measured density moments of the multimode object. The only restriction we have applied to the measurement operators $\mo{L}_i$ is the order of the field operators which we note are `number like' algebraically. This form is the lowest order number-conserving interaction possible for a multimode system.  Measurements that are lower order with respect to the field operators will not conserve number and may be suited to traditional coherent-state representations, but in many cases these systems may be treated using analytic  techniques like the Kalman filter, making simulation less important.  Number-conserving measurements are quite common in engineered monitoring of BECs \cite{Dalvit:2002,Szigeti:2009,Szigeti:2010,Thomsen:2002, Ruostekoski:1997, Corney:1998, Li:1998, Leonhardt:1999}, as any phase-sensitive measurement requires the existence of an atomic local oscillator to use as a phase standard.
Thus, the efficient simulation of Eq.~(\ref{eqn:mastmulti}) will be relevant to a wide variety of atom-optic systems, including all those involving current experimental detection schemes. 

To compare the performance of the NPW representation to coherent methods when simulating conditional master equations of the form Eq.~(\ref{eqn:mastmulti}) we require a verifiable solution for comparison. Unfortunately, Gaussian analytic techniques commonly applied in quantum control are not appropriate for Eq.~(\ref{eqn:mastmulti}). This can be understood by noting the measurement operator $\mo{L}_i$ is second order with respect to the field operators, which generates non-quadratic terms. Thus Gaussian analytic techniques such as the Kalman filter are not guaranteed to be exact \cite{Wiseman:2010}, and we are forced to integrate the master equation directly to generate a benchmark for comparison. This restricts us to looking at single mode systems, as direct integration is not scalable to multimode systems. The single mode problem that is algebraically equivalent to the multimode Eq.~(\ref{eqn:mastmulti}) is  
\begin{equation}
d\mo{\rho} = \gamma \mathcal{D}[\mo{a}^\dag\mo{a}] \mo{\rho} \; dt +  \gamma \mathcal{C}[\mo{a}^\dag\mo{a}] \mo{\rho} \; dt + \sqrt{\gamma} \mathcal{H}[\mo{a}^\dag \mo{a}] \circ dW. \label{eqn:mastsinglestrat}
\end{equation}
where $\circ dW$ is a Stratonovich Wiener increment and $\mathcal{C}[c] \rho = -\frac{1}{2} \left( \mo{c}^2 \rho + 2\mo{c} \rho \mo{c}^\dag + \rho (\mo{c}^\dag)^2 - \Tr[\mo{c}^2 \rho + 2\mo{c} \mo{\rho} \mo{c}^\dag + \mo{\rho} (\mo{c}^\dag)^2] \rho \right) + \left(\mo{c} \mo{\rho} + \mo{\rho} \mo{c}^\dag - \Tr[\mo{c} \mo{\rho} + \mo{\rho} \mo{c}^\dag ] \mo{\rho} \right)\Tr[\mo{c} \mo{\rho} + \mo{\rho} \mo{c}^\dag ]$ is the Stratonovich correction superoperator. This master equation is of a system undergoing continuous collopse under a number measurement.

The scalability of stochastic techniques for solving conditional quantum dynamics has already been demonstrated in \cite{Hush:2009}, but we aim to investigate the effect of choosing different representations. We use master equation Eq.~(\ref{eqn:mastsinglestrat}) to compare the performance of leading coherent-based scalable stochastic methods to the NPW representation. The convergence of these techniques is compared to a direct integration of the master equation.

We start our analysis with the coherent state based representations P$^+$ and TW. The success of these techniques have been primarily concerned with Hamiltonian and decoherence evolution of BEC and quantum-optical systems. Starting with P$^+$, we now investigate the applicability of these techniques on a conditional master equation. Using the correspondences in \cite{Gardiner:1991} we can convert the master equation (\ref{eqn:mastsinglestrat}) to
\begin{eqnarray}
d\mathcal{P}(\bm{\alpha}) & = & \left\{ \gamma \left[ \partial_{\alpha} \alpha \left( 1+ 2|\alpha|^2 - 2\sEq{\mathcal{P}}{|\alpha|^2}\right) \right. \right. \nonumber \\ 
& & + \partial_{\alpha^*} \alpha^*\left( 1+ 2|\alpha|^2 - 2\sEq{\mathcal{P}}{|\alpha|^2}\right) \nonumber \\ 
& & - \partial_{\alpha}^2 \alpha - \partial_{\alpha^*}^2 (\alpha^*)^2 \nonumber \\
& & - 2\left(|\alpha|^2 + |\alpha|^4 - \sEq{\mathcal{P}}{|\alpha|^2} - \sEq{\mathcal{P}}{|\alpha|^4}  \right) \nonumber \\
& & \left. + 4\sEq{\mathcal{P}}{|\alpha|^2}\left( |\alpha|^2 - \sEq{\mathcal{P}}{|\alpha|^2}\right) \right] dt \nonumber \\
& & + \sqrt{\gamma} \left[ - \partial_{\alpha} \alpha - \partial_{\alpha^*} \alpha^* \right.\nonumber \\
& & \left. \left. + 2\left(|\alpha|^2 - \sEq{\mathcal{P}}{ |\alpha|^2}\right) \right] \circ dW \right\} \mathcal{P}(\bm{\alpha}),  
\label{eqn:pprobstrat}
\end{eqnarray}
where $\mathcal{P}(\bm{\alpha})$ is the P-representation quasi-probability distribution which reproduces normally ordered moments of the master equation (\ref{eqn:mastsinglestrat}). Here $\sEq{\mathcal{Q}}{f(\mv{x})} \equiv \int d\mv{x}' f(\mv{x}') \mathcal{Q}(\mv{x}')$ is our notation for taking the expectation values of a function $f(\mv{x})$ with respect to the quasi-probability distribution $\mathcal{Q}(\mv{x})$.We immediately note that this equation contains non-positive definite diffusion that must be simulated by doubling the phase space. Thus P$^+$ techniques are required. This representation can then be unravelled into the following set of stochastic equations:
\begin{equation}
\begin{split}
d\alpha & = -2\gamma \alpha \left( \beta\alpha - \sEf{\beta \alpha} \right) dt \\
&\quad+ \sqrt{\gamma} \alpha \circ (i dV_1 + i dV_2 + dW);  \\
d\beta & = -2\gamma \beta \left( \beta\alpha - \sEf{\beta \alpha} \right) dt \\
&\quad+ \sqrt{\gamma} \beta \circ (-i dV_1 + i dV_2 + dW); \\
d\omega & = -2\gamma \omega \left( \beta \alpha + \beta^2 \alpha^2 - 2 \beta \alpha \sEf{\beta \alpha} \right) \\
&\quad+ 2\sqrt{\gamma} \omega \beta \alpha \circ dW. 
\end{split}
\label{eqn:psdesstrat}
\end{equation}
Where $dV_1$ and $dV_2$ are the set of `fictitious noises' that are averaged over to obtain the weighted averages $\sEf{f(\mv{x})} \equiv \sum_i \omega_i f(\mv{x}_i) / \sum_i \omega_i $. For more details on the techniques used to unravel equation (\ref{eqn:pprobstrat}) into (\ref{eqn:psdesstrat}) and how to simulate them see~\cite{Hush:2009}. Eqs.~(\ref{eqn:psdesstrat}) will be used to benchmark the unravelling of Eq.~(\ref{eqn:mastsinglestrat}) using coherent-based methods. 

We continue our analysis with the TW representation. Using the operator correspondences given in~\cite{Gardiner:1991}, we can write the master equation for the Wigner quasi-probability distribution $\mathcal{W}(\bm{\alpha})$ as
\begin{eqnarray}
d {\mathcal W}(\bm{\alpha}) & = & \left\{ \gamma \left[ - \frac{1}{8} \partial_{\alpha^*}^2 \partial_{\alpha}^2  -\frac{1}{2} \partial_{\alpha} ^2 \alpha^2 - \frac{1}{2} \partial_{\alpha^*}^2 (\alpha^*)^2 \right. \right. \nonumber \\
&+ &  \partial_{\alpha^*} \partial_{\alpha} \left(2 |\alpha|^2 - \sEq{\mathcal{W}}{|\alpha|^2}\right) \nonumber  \\
&- &  2\left( |\alpha|^4 - \sEq{\mathcal{W}}{|\alpha|^4} \right) \nonumber \\
&+ & \left.  4\sEq{\mathcal{W}}{|\alpha|^2} \left( |\alpha|^2 - \sEq{W}{|\alpha|^2}\right) \right] dt \nonumber \\
&+ & \left.  \left[ -\frac{1}{2} \partial_{\alpha^*} \partial_{\alpha} + 2\left(|\alpha|^2 - \sEq{\mathcal{W}}{|\alpha|^2}\right) \right] \circ dW \right\} \mathcal{W}(\bm{\alpha}). \nonumber \\
\label{eqn:wprobstrat}
\end{eqnarray}
Note that the first term contains higher order derivatives and a truncation is required in order to obtain a stochastic unravelling of Eq.~(\ref{eqn:wprobstrat}). Note also that a P$^+$ style extension of the phase space would be required to simulate the diffusion in the conditioning term. Traditionally the Wigner representation is guaranteed to produce strictly positive-definite diffusion~\cite{Gardiner:1991}, but this is under the assumption that the calculus increment is positive as is the case with $dt$. Unfortunately this assumption does not hold with the $dW$ increment.  A new `positive' Wigner representation could be derived by analogy to the P$^+$ representation, but the higher order terms would still need to be truncated.  This would make this hypothetical representation both approximate \textit{and} doubled in phase space, which would make it unlikely to compete with P$^+$. Thus it is not worthy of further investigation.

Finally we consider the number-phase Wigner representation. The NPW was first derived in \cite{Hush:2010} and was used in the simulation of large nonlinear equations. We now consider its applicability for use on conditioned large atom-optic systems.  Using the operator correspondences given in~\cite{Hush:2010} we get the following equation
\begin{eqnarray}
d{\mathcal N}(n,\phi) & = & \left\{ \gamma \left[ \frac{1}{2} \partial_{\phi}^2  - 2(n^2 - \sEq{\mathcal{N}}{n^2}) \right. \right.  \nonumber \\
& & \Bigl. + 4\sEq{\mathcal{N}}{n}(n - \sEq{\mathcal{N}}{n}) \Bigr] dt \nonumber \\
& & \Bigl. + 2 \sqrt{\gamma} \left[ n - \sEq{\mathcal{N}}{n} \right] \circ dW \Bigr\} {\mathcal N}(n,\phi), 
\label{eqn:nprobstrat}
\end{eqnarray}
where $\mathcal{N}(n,\phi)$ is the NPW representation that produces a complete set of moments of the master equation as outlined in~\cite{Hush:2010}. We next unravel Eq.~(\ref{eqn:nprobstrat}) using \cite{Hush:2009} to
\begin{eqnarray}
dn & = & 0; \nonumber \\
d\phi & = & \sqrt{ \gamma} dV_1; \nonumber \\
d\omega & = & \gamma \omega (- 2n^2 + 4\sEf{n} n) dt + \sqrt{\gamma} \omega n \circ dW.  \label{eqn:nsdesstrat}
\end{eqnarray}
Note we did not need to apply any truncations or double the phase space. The simplicity of the equations (\ref{eqn:nsdesstrat}) compared to (\ref{eqn:psdesstrat}) show how an appropriate choice of representation, NPW in this case, can greatly reduce the complexity of the evolution, just as an appropriate choice of basis can simplify analysis of other quantum problems. We can now compare the performance of the NPW representation to P$^+$ by integrating equations (\ref{eqn:psdesstrat}) and (\ref{eqn:nsdesstrat}), respectively.

\begin{figure}[htb]
\includegraphics[scale=0.52]{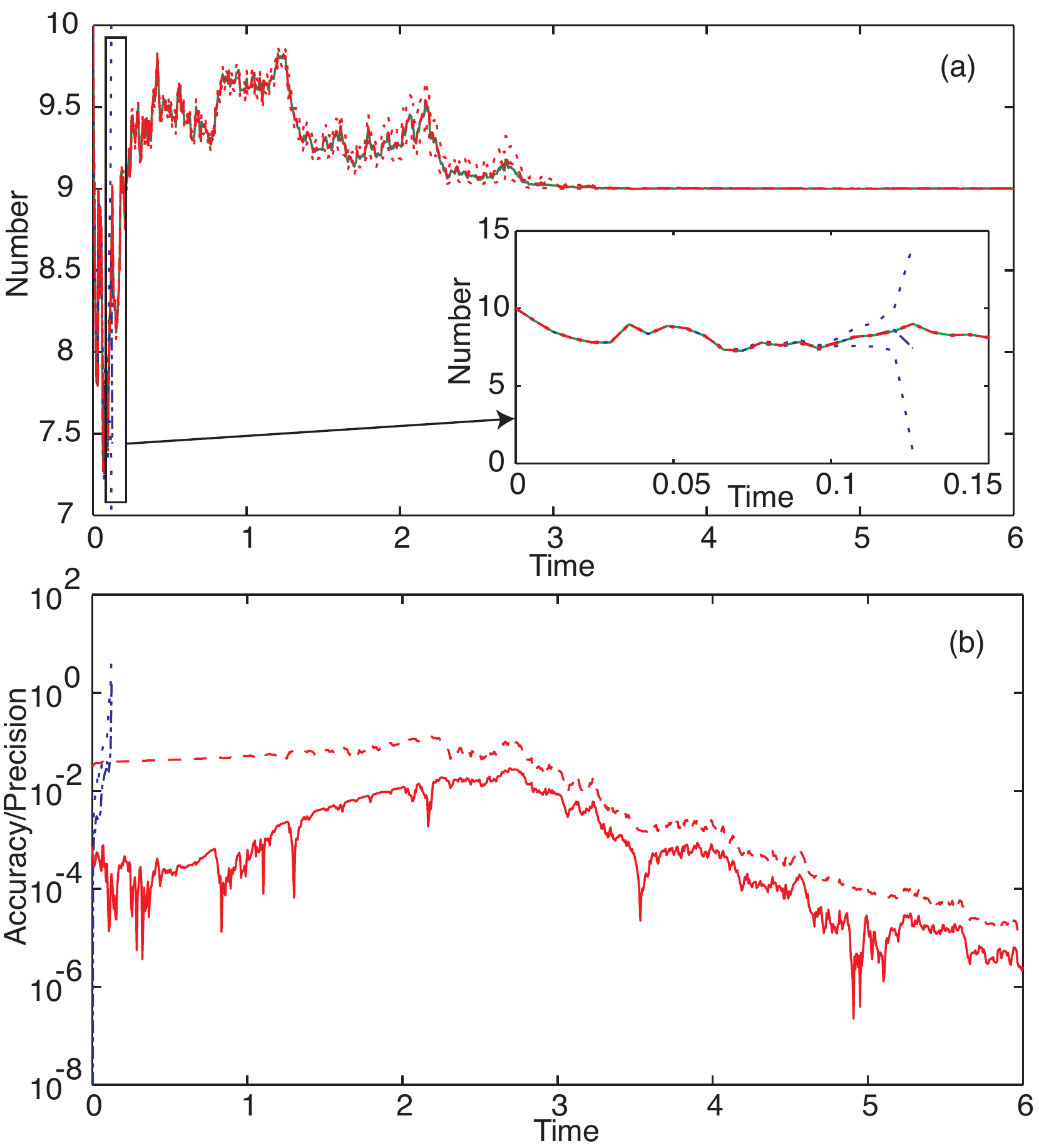}
\caption{(Color online) (a) Plot of number versus time (in dimensionless units) for a single measurement run, comparing the direct integration of the master equation to the NPW and P$^+$ methods.  Simulations had an initial condition of a coherent state with an amplitude of 10.  Numerical integration of the master equation (\ref{eqn:mastsinglestrat}) is plotted with a solid green line, the NPW (\ref{eqn:nsdesstrat}) is plotted with a dashed red line and P$^+$ (\ref{eqn:psdesstrat}) is plotted with a dot-dashed blue line.  Uncertainty in both stochastic methods is plotted with dotted lines. The P$^+$ becomes divergent around $t=0.15$, and is not plotted beyond this point. A close-up view of this divergence is shown in the inset. Part (b) shows the accuracy, defined as the difference between each stochastic result and the master equation solution, and the precision, defined as the standard deviation of the averages, for each stochastic method on a logarithmic scale.  The NPW accuracy is described with a solid red line, the NPW precision is a dashed red line, the P$^+$ accuracy is a dash-dotted  blue line, and the P$^+$ precision is a dotted blue line.  The NPW is considerably more precise and is convergent for at least an order of magnitude longer than the competing P$^+$ representation. The numerical integration was performed by using the open source software package XMDS \cite{xmds:0000}.}
\label{fig:comp}
\end{figure}

The simulations comparing the P$^+$, NPW and a direct integration of the master equation  are shown in Fig.~\ref{fig:comp}.  The number-phase Wigner representation converges for the longest time interval. In fact it converges until a complete collapse into the correct number state. As this is the steady state of the equation we expect the number-phase representation to converge indefinitely. Not only is the NPW more accurate it is also significantly more precise. The increase in precision in turn improves the accuracy of the evolution, as simulation of the conditional master equation uses an estimate of the observable $\Tr[\mo{c} \mo{\rho} + \mo{\rho} \mo{c}^\dag]$, thus lack of precision results in a lack of accuracy in the long term. This dynamic instability is not seen in non-conditional master equation evolution, and makes the precision of stochastic techniques considerably more important in these problems.  

The results show that the NPW-based simulations are significantly better than the simulations based on coherent-state representations for the conditional master equation described in Eq.~(\ref{eqn:mastsinglestrat}). The simulations promise to be stable enough to consider long-term behaviour of systems and model the effects of feedback strategies.  This high level of convergence can be explained by noting the suitability of the basis underlying the representation to the measurement eigenstates.  Importantly, due to practical difficulties in producing stable atomic local oscillators, any current detection scheme used in quantum gases is of a form that is suited to the NPW method.  The NPW method described in this letter is also the only simulation tool that is deterministic for the strong number-conserving nonlinearities that are present in such systems, and is therefore the only suitable candidate for simulating conditional states of ultracold atomic gases for feedback or state estimation. 

This research was supported under Australian Research Council's Discovery Projects funding scheme (project number DP0556073) and the Australian Research Council Centre of Excellence for Quantum-Atom Optics (ACQAO). We acknowledge the use of CPU time at the National Computational Infrastructure National Facility and thank Graham Dennis for his help with simulations.

\bibliography{papers}

\end{document}